**Dirac Fermion Quantum Hall Antidot in Graphene**

Scott Mills[†], Anna Gura[†], Kenji Watanabe[‡], Takashi Taniguchi[‡], Matthew Dawber[†], Dmitri Averin[†*] and Xu Du[†*]

[†]Department of Physics and Astronomy, Stony Brook University, Stony Brook NY 11794-3800, USA

[‡]National Institute for Materials Science, 1-1 Namiki, Tsukuba 305-0044, Japan

**Abstract**

The ability to localize and manipulate individual quasiparticles in mesoscopic structures is critical in experimental studies of quantum mechanics and thermodynamics, and in potential quantum information devices, e.g., for topological schemes of quantum computation. In strong magnetic field, the quantum Hall edge modes can be confined around the circumference of a small antidot, forming discrete energy levels that have a unique ability to localize fractionally charged quasiparticles. Here, we demonstrate a Dirac fermion quantum Hall antidot in a graphene, where charge transport characteristics can be adjusted through the coupling strength between the contacts and the antidot, from Coulomb blockade dominated tunneling under weak coupling to the effectively non-interacting resonant tunneling under strong coupling. Both regimes are characterized by single -flux and -charge oscillations in conductance persisting up to temperatures over 2 orders of magnitude higher than previous reports in other material systems.  Such graphene quantum Hall antidots may serve as a promising platform for building and studying novel quantum circuits for quantum simulation and computation.

**\*** email: dmitri.averin@stonybrook.edu, xu.du@stonybrook.edu

## I. Introduction

Localization and manipulation of individual quasiparticles play an important role in studies of quantum mechanics and quantum thermodynamics, and in the applications of quantum information devices. A wide variety of mesoscopic systems, such as quantum dots[1-4], nitrogen-vacancy (NV) centers[5, 6], superconducting Cooper pair boxes[7-10], superconducting quantum interference devices[11], etc., have been extensively studied for quantum manipulation of charges, spins, and magnetic fluxes. Quantum Hall (QH) antidots, on the other hand, offer a promising approach for localizing quantum Hall quasiparticles. Due to quantum confinement, the chiral one-dimensional (1D) edge mode has its energy quantized into discrete levels, mimicking a large, tunable "artificial atom" which hosts QH quasiparticles. Compared to the other approaches, QH antidots are capable of localizing even exotic quasiparticles with fractional charges and nontrivial exchange statistics. It therefore holds promise for topological schemes of quantum computation[12, 13]. Experimentally, pioneering studies of QH antidots have been carried out using GaAs two-dimensional electron gas (2DEG), where localization of integer[14, 15] and fractionally charged[16] quasiparticles have been demonstrated. The potential of applying such QH antidots for quantum information applications has also been discussed[17]. On the other hand, due to the small energy-scale associated with Landau level (LL) spacing and energy quantization, the signatures of the localized QH edge states, namely charge and magnetic flux oscillations in conductance across the antidot, are fragile and require very low electron temperature (typically sub-100 milliKelvin)[18] to observe. A 2DEG system which can provide more robust localization of QH quasiparticles and stronger coherence is therefore desirable for realizing more complex devices and functionalities.

The development of 2D crystal graphene in the recent decade raised a new opportunity for studying localized QH states in the antidot setup. The Dirac-nature of the 2DEG in graphene[19] differs fundamentally from that in GaAs, due to its linear energy dispersion, chirality and non-trivial Berry's phase. As a result, graphene can achieve high charge carrier mobilities ($> 10^5$ cm$^2$/Vs) which persist even with densities down to $\sim 10^9$ cm$^{-2}$ without localization. It has a large and energy-independent Fermi velocity ($v_F \approx 10^6$ m/s), which leads to large LL spacing ($\Delta\varepsilon_{LL} = \sqrt{2e\hbar v_F^2 B} = 35\sqrt{B[\text{T}]}$ meV)[20], as well as large quantization energy spacing under confinement. Both are critical factors for realizing robust localization of QH states. Technically, being a single atomic layer, the size of graphene devices may be pushed down to nanometer-scale with sharp definition[21]. And Ohmic electrical contacts with low contact resistance have been routinely achieved both for top contacts[22] and side contacts[23]. Despite all these promising characteristics, experimental work on confining QH edge states in monolayer graphene has been limited, and mainly focused on gate-defined quantum dots[24, 25]. Charge transport studies on well-structured graphene-based QH antidots, on the other hand, have not been reported to our knowledge. In this work, we study Dirac electron QH antidots in graphene, and demonstrate robust localization of QH edge states in the lowest -LL (LLL) which persists up to 10-100 times higher temperature compared to previous reports.

## II. Results and Discussion

The samples used in this work, illustrated in Figure 1, are point contact-coupled antidots embedded in hexagonal boron nitride (hBN)-encapsulated graphene field effect transistors. The two-terminal conductance consists of a background from the bulk[26] and the tunneling conductance through the QH antidot. Under well-developed QH effect, depending on the coupling strength between the point contacts and the antidot, the QH plateau resistances can vary between $\frac{h}{\nu e^2}$ for decoupled antidot, and $\frac{h}{2\nu e^2}$ in the situation when antidot is split by the contacts (see Supplementary Information). In our experiments, the coupling between the point contacts and the antidot is carefully adjusted to be in between these two limits. The overall geometry of our devices is designed so that: 1) the diameter of the antidot $D_{dot} \gg l_B = \sqrt{\frac{\hbar}{eB}}$, where $l_B$ is the magnetic length ~10nm for the few-Tesla magnetic field applied here; 2) $D_{dot}$ is sufficiently small for energy quantization; 3) the point contacts probing the antidot are sufficiently sharp to minimize invasive effects and to optimize phase coherent charge transport. Satisfying these criteria, we designed the diameter to be $D_{dot} \sim 200 - 310$nm and the width of the point contacts to be ~ 15nm. The dimensions of the devices are directly confirmed by SEM imaging. Figure 1C shows QH resistance as a function of filling factor $\nu = \frac{nh}{eB}$ in various magnetic fields (through ramping of the gate voltage). We note that in measuring these curves the gate voltage ramping speed is relatively large and the fine oscillatory features (as discussed below) are washed out. At anomalous integer fillings $\nu = 4\left(n + \frac{1}{2}\right)$, the resistance values are observed to be between $\frac{h}{\nu e^2}$ and $\frac{h}{2\nu e^2}$ due to the conduction through the antidot.

In quantizing magnetic fields, periodic conductance oscillations both in gate voltage and in magnetic field become prevalent at filling factors between $\nu = 1$ and $\nu = 2$, with typical amplitude of ~$0.1 G_0$ ($G_0 = \frac{e^2}{h}$ is the conductance quantum). Depending on the coupling distance between the point contacts and the antidot, two types of periodicity are observed which we separately discuss below. In the case of relatively weak coupling, as shown in Figure 1B for a 310nm-diameter antidot sample with a few tens of nanometers contact-antidot separation, the conductance oscillations under ramping magnetic field show an approximately constant period of

21 mT. In Figure 2B, we plot the oscillatory conductance as a function of magnetic flux through the antidot, normalized over single electron magnetic flux quantum: $\frac{\Phi}{\Phi_0} = \frac{eB\pi D_{QH}^2}{4h}$, where $D_{QH}$ is the diameter of the QH edge current encircling the antidot, and $\Phi_0 = \frac{h}{e}$. With $D_{QH} = 350 nm$ which is slightly larger than the physical diameter of the antidot $D_{dot} = 310 nm$ (by the amount consistent roughly with the notion that the edge states are removed from the physical edge by the distance of an order of the magnetic length $l_B$), the observed conductance oscillations match closely with a flux period of $0.5 \pm 0.1 \Phi_0$. We note that the observed $\frac{1}{2}\Phi_0$ flux oscillations are reproducible over repeated thermal cycles and over different samples with similar weak coupling. This suggests their origin to be intrinsic to the antidots, and independent of the unintentional defects.

Corresponding periodic conductance oscillations are also observed as the gate voltage is swept at fixed magnetic field. Figure 2D shows the conductance oscillations in the 310 nm antidot as a function of gate voltage and charge number change over the area of the antidot: $\frac{\Delta Q}{e} = \frac{c\Delta V_G \pi D_{dot}^2}{4e}$. Here $c$ is the geometrical capacitance per area. On average, each conductance oscillation corresponds to addition/removal of one electron into the edge state on the antidot. Compared to the flux oscillations, the charge oscillations appear to be somewhat less regular indicating the presence of random fluctuations in the coupling capacitance between the back gate and the antidot. Direct correspondence between the charge and flux oscillations is evidenced by Figure 2E, where the conductance oscillations form tilted stripes on the "magnetic field-gate voltage" plane. Along the charge oscillation axis, however, there are significant random fluctuations which manifest the charge noise.

Next, we discuss the antidots with stronger coupling to the point contacts, where a different type of magnetic oscillation periodicity is observed, as illustrated by Figure 3A for a 200nm-diameter antidot with point contacts touching its circumference (SEM image shown in the Supplementary Information). Here, with increasing magnetic field, the conductance oscillations evolve from single-peaks (Figure 3A inset) to "M"-shaped double-peaks (Figure 3A main panel). With an estimated $D_{QH} \sim$ 225nm (which again differs from the physical diameter by a size of the order of magnetic length), both the single-peak oscillations in low field and the "M"-shaped pairs in stronger field match with a flux period of $\phi_0$. The flux separation of the two conductance peaks within an "M"-shaped pair ($\Delta B_Z$) increases with increasing magnetic field, as discussed in detail later. Corresponding to the magnetic oscillations, the gate voltage dependent conductance oscillations in this 200nm-diameter antidot also show "M"-shaped pairs, as plotted in Figure 3B. The charge period in in this sample is within ~20% of what is calculated using the geometric size of the antidot: $\frac{\Delta Q}{e} = \frac{c\Delta V_G \pi D_{dot}^2}{4e}$. The discrepancy may be attributed to the errors in the estimations of the antidot size and effective gate capacitance at the antidot.

The observation of $\Phi_0$ period (which obviously also includes the $\frac{1}{2}\Phi_0$ period) of the magnetic flux oscillations strongly suggests Aharonov-Bohm (AB) effect. We note that both charge and flux oscillations are maximized when the Fermi energy is between the zeroth LL and the first LL, and that the oscillations disappear during the plateau-to-plateau transition when the Fermi energy coincides with one of the LLs. This suggests that the oscillations are associated with the QH edge states encircling the antidot. Indeed, clear AB oscillations can be present only if the circulating current has a well-defined diameter. In a QH system, this happens when the Fermi energy is between the LL, and, with the bulk gapped, only the quasi-1D edge modes conduct.

In a single-particle picture (at first, neglecting Zeeman effect), the energy quantization of the finite-size QH edge state is obtained from the single-particle Dirac Hamiltonian $H_0 = v_F \vec{\sigma} \cdot (\vec{p} - e\vec{A}) + U$ with boundary conditions imposed by the antidot geometry (see Supplementary Information). Here $\vec{A} = rB\hat{\varphi}$ is the vector potential in symmetric gauge. The potential energy $U$ is constant in the single-particle picture. Considering the relevant length scales: $D_{dot} \gg l_B \gg a$ (where $a \sim 0.14$ nm is the lattice constant), the edge state can be approximated as encircling the antidot edge with diameter $D_{QH} \approx D_{dot}$. The Dirac Hamiltonian results in magnetic field-dependent energy levels $\varepsilon_j = \frac{2j\hbar v}{D_{QH}} + \frac{ev\Phi}{\pi D_{QH}}$, with constant level spacing (neglecting disorder) $\delta\varepsilon = 2\hbar v/D_{QH}$. Here $v = -\frac{1}{eB}\frac{dV}{dr} \lesssim v_F$ ($V$ being the confinement potential at the edge) is the velocity of the QH edge state. We note that this relation gives the proper account of the AB effect, as the periodicity of the antidot energy spectrum in magnetic flux has a period equal to the magnetic flux quantum: $\varepsilon_j(\Phi) = \varepsilon_{j-1}(\Phi + \Phi_0)$. And in graphene, the sharp definition of the antidot facilitates large energy level spacing $\delta\varepsilon$.

Electron-electron interaction leads to Coulomb-blockade-type effects. In the weak coupling limit ("closed" antidot), the electron interaction energy $U_{ee} = \frac{e^2}{2\pi\epsilon\epsilon_0 L}\ln(\frac{L}{l_B})$, where $L = \pi D_{dot}$ is the circumference of the edge, is constant for the relevant edge states (see Supplementary Information). For the QH antidots studied in this work, the interaction energy is $U_{ee} = 2.3 - 3.2$ meV, which is much smaller than the LL spacing but significantly larger than the Zeeman energy within the range of magnetic fields applied. In the presence of metallic contacts, $U_{ee}$ should be screened and somewhat reduced.

Following the discussion above, the antidot can be described by a many-body Hamiltonian (see Supplementary Information):

$$H = \frac{U_{ee}}{2}(n + n_\phi - n_G)^2 + \sum_{j,\sigma} \varepsilon_{j,\sigma} n_{j,\sigma}$$

Here $n = \sum_{j,\sigma} n_{j,\sigma}$ is the total charge number associated with the edge states; $n_{j,\sigma}$ is the occupation number of the state with orbital index $j$, spin index $\sigma = \pm 1$, and the single-particle energy $\varepsilon_{j,\sigma} = \varepsilon_j + g\mu_B B\sigma/2$, where $\mu_B$ is the Bohr magneton, $B$ - the magnetic field, and the g-factor for electrons in graphene is close to its free-electron value[27] $g \approx 2$. Also in the Hamiltonian, $n_G = cV_G$ and $n_\phi = \frac{\nu_S \Phi}{\Phi_0}$, where $\nu_S$ is the number of occupied LLs including spin degeneracy which gives the number of the propagating edge modes. Periodicity of the linear antidot conductance, or other equilibrium properties of the antidot, in the back-gate voltage $V_G$ or magnetic flux $\Phi$ is determined by the behavior of the occupation factors $n_{j,\sigma}$ and the Hamiltonian $H$ as functions of $V_G$ and $\Phi$ under the conditions of the fixed chemical potential of the external contacts to the antidot. Increase of the gate voltage such that $n_G \rightarrow n_G + 1$ leads to the corresponding increase of the total occupation factor, $n \rightarrow n + 1$, leaving the Hamiltonian invariant, $H(n_G) = H(n_G + 1)$. Similarly, because the spin-degenerate single-particle spectrum of the antidot is periodic in flux with the period $\Phi_0$, increase of the flux such that $n_\phi \rightarrow n_\phi + \nu_S$ leads to the decrease of the total occupation factor, $n \rightarrow n - \nu_S$, leaving the Hamiltonian unchanged, $H(n_\phi) = H(n_\phi + \nu_S)$. This gives rise to the observed single-charge periodicity in the gate voltage, and ensures $\Phi_0$ flux periodicity in all regimes in agreement with the general Byers-Yang theorem[28]. In addition, in Coulomb interaction-dominated regime, the Hamiltonian also gives rise to finer flux oscillations with a periodicity of $\frac{\Phi_0}{\nu_S}$, which corresponds to $\Delta n_\phi = 1$.

For a relatively weakly coupled ("closed") antidot (e.g., the 310 nm-diameter antidot sample in Figure 1B) and with the Fermi energy above the LLL ($v_S = 2$), equally spaced magnetic oscillations are observed with $\Phi_0/2$ period (consistent with the larger period $\Phi_0$) independent of magnetic field. This suggests that the antidot tunneling is dominated by electron-electron interaction which results in a constant energy gap for the addition of individual electrons to the antidot that is independent, e.g., of electron spin. To find this addition energy, we measured the bias voltage dependence of the device conductance. In Figure 2F, we plot the background-normalized differential conductance (*dI/dV*) as a function of gate voltage and bias voltage (charge stability diagram), which resembles the Coulomb blockade diamonds, revealing the charge addition energy to be ~1.5meV for the 310nm diameter antidot sample. We note that the quality of the charge stability diagram is limited by the charge noise in the sample and the electrical noise in our measurement setup. Well defined periodicity of the conductance oscillations in this antidot implies that the addition energy is given by the Coulomb repulsion energy $U_{ee}$, with the single-particle energy level spacing $\sim \delta \varepsilon$ either too small to be noticeable in comparison to $U_{ee}$, or washed out by electron-electron relaxation. Compared to a simple "closed-antidot" estimate of ~2.3meV, the observed Coulomb repulsion energy is somewhat smaller, which can be qualitatively explained by finite screening effect from the contacts.

For a strongly coupled ("open") antidot, the electron-electron interaction is largely screened by the electrodes. Neglecting electron correlations, the single-particle picture of energy level quantization and Zeeman splitting predicts AB oscillation with a primary oscillation period of $\Phi_0$ and a B-dependent Zeeman splitting of the conductance peaks in strong magnetic fields. This is indeed consistent with our observation where single-flux oscillations in weak magnetic field split and resolve into "M"-shape pairs in strong magnetic field. In Figure 3C, the conductance oscillation period in magnetic field is plotted as a function of magnetic field, both for the "M"-shaped pairs ($\Delta B$) and for the magnetic field splitting within each pair ($\Delta B_Z$). $\Delta B$ is roughly magnetic field independent with its corresponding flux period $\sim \Phi_0$. The magnetic field dependence of $\Delta B_Z$ can be fit to a straight line intersecting the origin. Extrapolating the linear dependence, $\Delta B_Z (B_0) \frac{\pi D_{QH}^2}{4} = \Phi_0$, we get $B_0 \sim 18.5$T at which Zeeman splitting becomes equal to the quantization energy spacing $\delta\varepsilon = 2\hbar v/D_{QH}$. Based on this comparison, we can estimate $\delta\varepsilon = g\mu_B B_0 \sim 2.1$meV. This value is confirmed by measuring the bias and gate dependence of the differential conductance (Figure 3D) where the level spacing energy is a sum of the heights of the two adjacent "diamonds" (from Zeeman splitting) in the plot, which gives ~ 2-2.5meV. From the level spacing, we can estimate the QH edge state velocity $v \sim 3.4 \times 10^5$m/s, about 1/3 of the Fermi velocity of free Dirac electrons in graphene.

In the Coulomb blockade-dominated regime, the conductance oscillations are suppressed at elevated temperatures through thermal excitation/smearing (Figure 4A). The corresponding antidot conductance can be calculated numerically through the linear response of the Coulomb blockade rate equations[29] to the external bias (see Supplementary Information). Figure 4B shows the temperature dependence of the averaged conductance oscillation amplitude (defined as the difference in conductance at successive minima and maxima), with the best fit calculated numerically with addition energy $U_{ee}$ and a normalization prefactor as fitting parameters. The best fit of the addition energy, $U_{ee} = 1.3$meV, is in good agreement with the value obtained experimentally from the height of the Coulomb diamonds for the 310 nm antidot sample (Figure 2F). As is characteristic of single electron devices, the linear conductance oscillations are nearly completely suppressed as the temperature exceeds half of the addition energy.

We note that in the weakly couple QH antidot discussed above, the Coulomb oscillations persist up to ~4K, which is 1-2 orders of magnitude higher than the previous reports for GaAs-based devices with comparable antidot size. The robustness of single charging effects can be further enhanced by increasing the electron-electron interaction energy $U_{ee}$. To demonstrate this, we study a QH antidot sample with ~250nm effective diameter made on suspended graphene, where $U_{ee}$ is enhanced through a reduction in dielectric screening. In the suspended graphene sample, carrier mobility can be sequentially improved through repeated current annealing (controlled Joule heating which evaporates the contaminants). Rather than coupling to the antidot using protruding point contacts (which induces unwanted high current density at the sharp tips, causing damage during the annealing), straight source and drain electrodes are used which are separated from the antidot by ~200nm (Figure 4D inset). Coupling to the antidot is possible only in low magnetic fields, through extended states from the electrodes. With high mobility > $10^5 cm^2/Vs$, QH plateaus become well developed in magnetic field as low as 1 Tesla. Figure 4D shows the conductance oscillations in gate voltage at B=1T. The antidot is weakly coupled as is evident from the nearly un-perturbed QH plateau resistance background (Figure 4D). A thorough current annealing narrows the conductance peaks, as shown by Figure 4E inset. In absence of the hBN encapsulation, the effective dielectric constant and screening becomes significantly reduced, resulting in a much larger Coulomb gap. In Figure 4F, the charge stability diagram measures a Coulomb gap of ~8meV. The large Coulomb gap allows single charge oscillations to persist for temperatures over 10K, as shown in Figure 4E. In Figure 4F, we also observe evidence of energy quantization in the excited states, indicated by the bright lines outside the Coulomb diamonds. The energy level spacing $\delta\varepsilon$ is found to be ~5meV, and is spin-degenerate due to the small magnetic field. Formation of such quantized energy levels suggests robust coherence in these high mobility samples.

For samples with strong coupling to the antidot, where Coulomb blockade is largely screened, both thermal excitation and disorder can contribute to the suppression of conductance oscillations (see Supplementary Information). Inelastic charge carrier scattering, including electron-electron and electron-phonon scattering, have a characteristic energy scale of ~ $k_B T$ which is much smaller than the energy level spacing $\delta\epsilon \approx 2\text{meV}$ for the 200nm antidot. Consequently, decoherence is not expected to play critical role in the temperature dependence of conductance oscillations. In the strong tunneling limit, the amplitude of the conductance oscillations can be calculated using a standard double barrier transmission model, characterized by the transmission probabilities at the two leads. In the LLL at fixed magnetic field, the two modes for each spin contribute in parallel with a relative shift given by the Zeeman energy. Increasing temperature in the two leads then effectively smears the energy levels and causes suppression of the amplitude of conductance oscillations (See Supplementary Information). In figure 3B we fit the gate-dependence of a "M"-shaped oscillation pair at $B = 8.5\text{T}$, using 0.85 and 0.51 as the transmission probabilities, and a Zeeman splitting of 0.82meV which is in reasonable agreement with $g\mu_B B = 0.98\text{meV}$. In Figure 4C the temperature dependence of the averaged conductance oscillation amplitude at $B = 10\text{T}$ is calculated using the thermal excitation model and compared with the experimental data, with transmission probabilities 0.72 and 0.41. The deviation of the simulation from the conductance oscillation amplitude data at low temperatures (T<1K) may be an indication of the charge noise in the vicinity of the antidot, which provides an additional mechanism of suppression of the conductance oscillations that becomes noticeable once the thermal broadening is weak.

## III. Conclusion

We demonstrate the first experimental study on Dirac electron quantum Hall antidots in graphene. Depending on the coupling strength to the antidot, both Coulomb blockade dominated tunneling and effectively non-interacting resonant tunneling are achieved. Both regimes are characterized by single-flux and single-charge oscillations in conductance which, due to the Dirac nature of the electron gas in graphene, persist up to temperatures over 2 orders of magnitude higher than that reported in previous reports on conventional 2D electron gas.

The main advantages of using graphene in QH antidots come from its large edge state velocity $v$ (as a result of the Dirac spectrum), which results in the large energy level spacing, and large Landau level separation which permits scaling up of the characteristics antidot energies. Another technical convenience of graphene in QH antidot samples comes from the observation that the diameter of the QH edge state encircling the antidot is very close to the physical size. All this opens the possibility of precise design of the antidots and their coupling into multiple antidot structures for possible applications in quantum information. We note that while in this work we mainly focused on direct metallic point contacts, coupling to the antidot can also be achieved through the more conventional QH edge-to-edge tunneling (see Supplementary Information). With further work to optimize the device structure and mobility, the graphene QH antidot system demonstrated here may serve as a promising platform for studying localized QH states, and for building antidot-based quantum circuits for quantum simulation and computation. Besides graphene, the chiral edge state-based antidot structure can also be applied in many other 2D systems (e.g., 2D superlattice, layered topological material, etc.), providing an effective technique for studying novel chiral quasiparticles.

**Acknowledgement**

This work was supported by AFOSR under grant FA9550-14-1-0405 and NSF under grant FP00000178. K.W. and T.T. acknowledge support from the Elemental Strategy Initiative conducted by the MEXT, Japan and the CREST (JPMJCR15F3), JST. S.M. and X.D. acknowledge Vladimir Goldman for insightful discussions.

**Figure Captions**

*Figure 1. Graphene QH antidot.* **A**. *Schematics of QH antidot. The red arrows indicate the QH edge states. The band structure illustrates the bulk LLs and the quantized edge states.* **B**. *Structure of a point contact-coupled QH antidot on graphene/hBN heterostructure. Inset: SEM image of a 310nm-diameter antidot sample. The yellow dotted lines highlight the boundaries of the antidot and the side contacts. The scale bar is 200nm. The bulk channel edges are outside the view.* **C**. *Two terminal resistance as a function of filling factor, measured in various fixed magnetic fields by ramping the back gate voltage.*

*Figure 2. Coulomb blockade in relatively weakly coupled QH antidot. Conductance oscillations are observed with ramping magnetic field (**A**) and gate voltage (**B**). (**C**)A zoom-in of the conductance oscillations shows a magnetic flux period of $\Phi_0/2$, calculated using a QH edge state diameter of $D_{QH}=350nm$.* **D**. *A zoom-in of the conductance oscillations shows approximate one oscillation per charge, with charge number calculated using the geometric capacitance over the physical area of the antidot.* **E**. *Conductance oscillations versus both gate voltage and magnetic field.* **F**. *Charge stability diagram showing Coulomb blockade characteristics. The dotted lines are guides to the eyes.*

*Figure 3. AB oscillations in strongly coupled QH antidot.* **A**. *Conductance oscillations showing $\Phi_0$ flux period and Zeeman splitting. Inset: In lower magnetic fields, the conductance oscillation evolves from merged peaks with $\Phi_0$ period to "M"-shaped spin-resolved oscillations, with increasing magnetic field.* **B**. *Conductance oscillations with ramping gate voltage. The top x-axis shows the change in charge number calculated using the geometrical capacitance over the physical area of the antidot. The red solid curve shows the fitting to a relatively symmetric "M"-shaped oscillation using thermal excitation model.* **C**. *Magnetic field dependences of the AB oscillation period ($\Delta B$) and normalized Zeeman splitting ($\Delta B_Z/ \Delta B$). The dotted lines are guides to the eyes.* **D**. *Differential conductance versus bias voltage and gate voltage. Dotted lines are guides to the eyes. From the bias- and gate-dependence of the differential conductance, the energy level spacing is a summation of the bias voltages at the two adjacent diamond tips which correspond to the two spin states.*

*Figure 4. Thermal suppression of conductance oscillations in the QH antidots.* **A**. *Gate dependent conductance oscillation at various temperatures. The curves, taken at (from top to bottom)T=0.4, 1.8, 2.5, 3.1, 3.8, 4.3, 4.8 and 5.5K, are vertically shifted for clarity. The temperature dependence of the averaged conductance oscillation amplitude for the weakly coupled 310nm antidot (**B**) and the strongly coupled 200nm antidot (**C**) are fitted with thermal excitation models.* **D**. *Conductance oscillations in a suspended graphene QH antidot at B=1T. The dotted line labels QH plateau resistance at n=2. Inset: SEM image of a suspended graphene with 3 QH antidot devices. The scale bar is 1μm* **E**. *Temperature dependence of conductance oscillations. The dotted and the solid black curves, both measured at 450mK, correspond to before and after thorough current annealing, respectively. The red, blue and purples curves are taken at 2K, 5K and 10K respectively, and are shifted downward for clarity.* **F**. *Charge stability plot for the suspended graphene QH antidot, taken in 1T magnetic field at 450mK. The dotted lines are guide to the eyes.*

**Figure 1.**

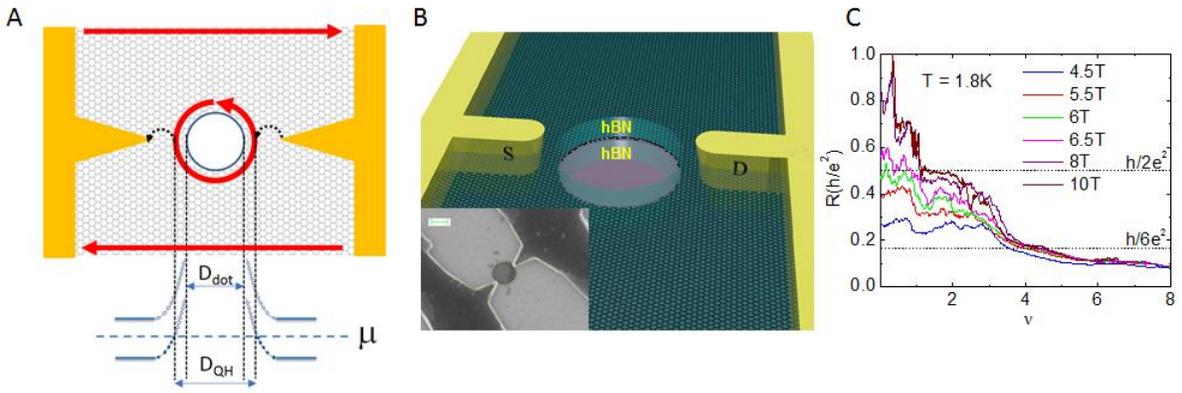

**Figure 2.**

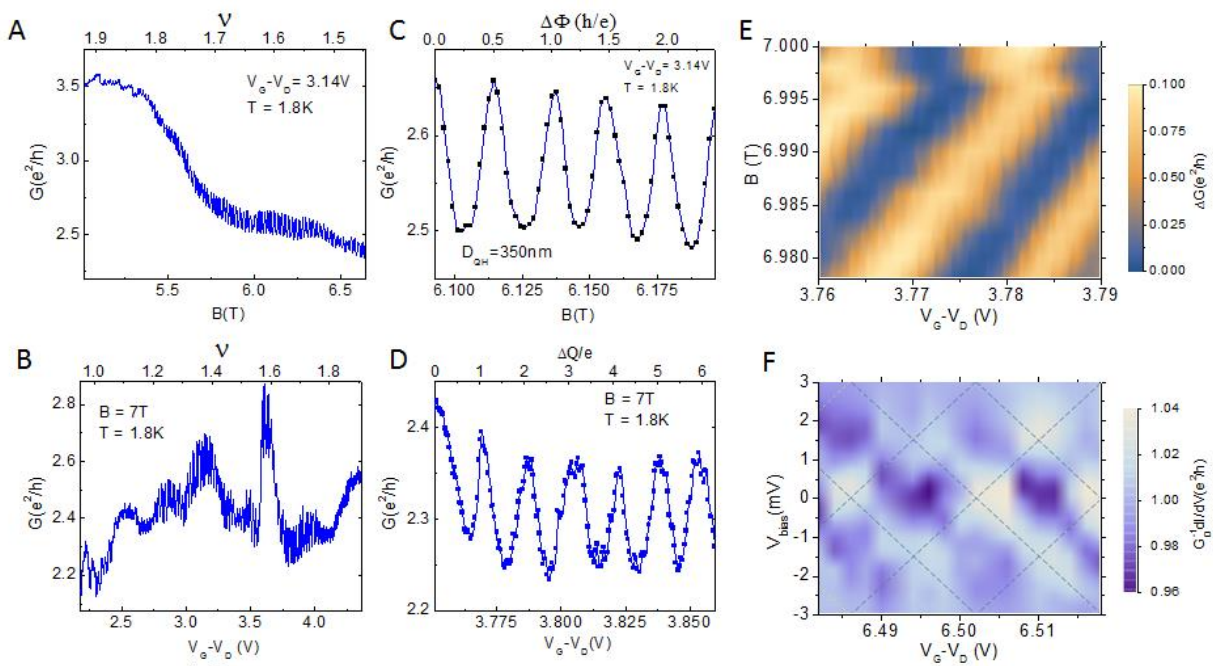

**Figure 3.**

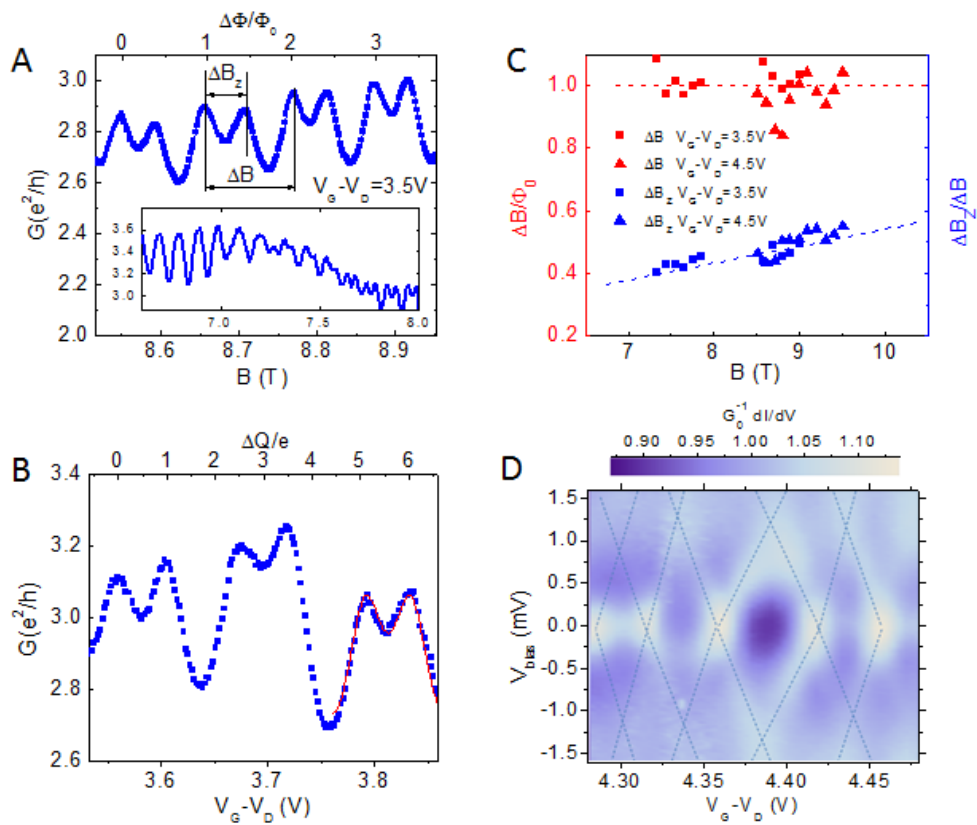

**Figure 4.**

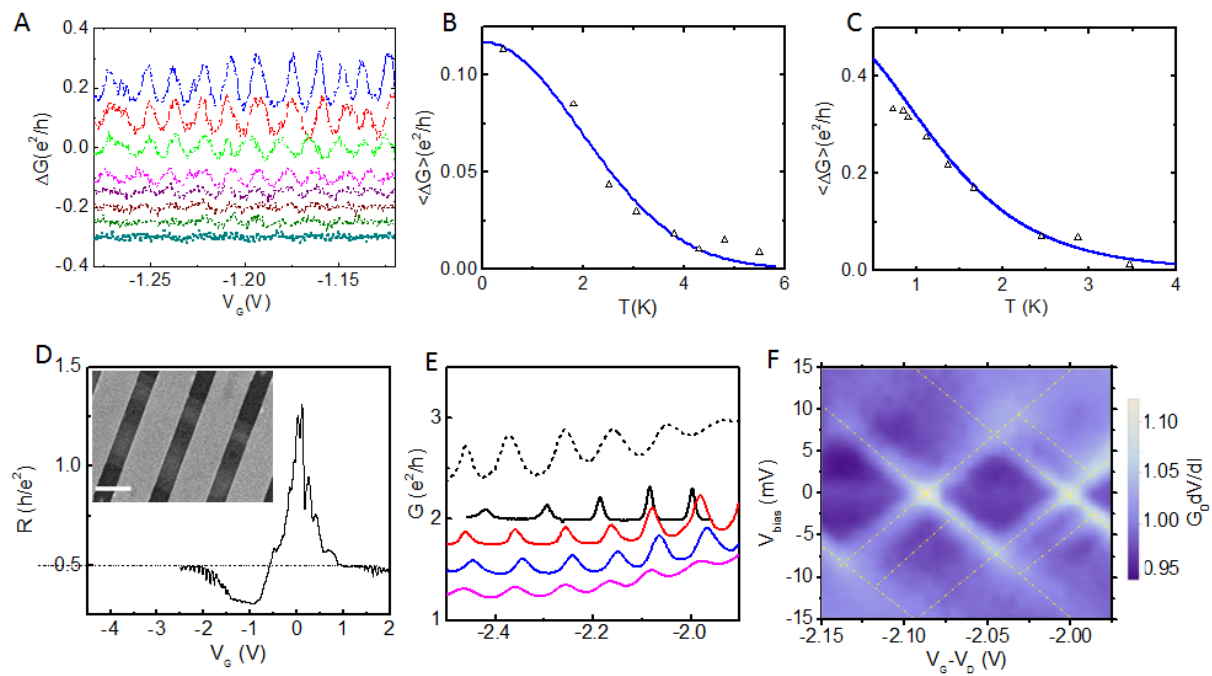

**Supplementary Information: Dirac Fermion Quantum Hall Antidot in Graphene**

Scott Mills[†], Anna Gura[†], Kenji Watanabe[‡], Takashi Taniguchi[‡], Matthew Dawber[†], Dmitri Averin[†*] and Xu Du[†*]

[†]Department of Physics and Astronomy, Stony Brook University, Stony Brook NY 11794-3800, USA

[‡]National Institute for Materials Science, 1-1 Namiki, Tsukuba 305-0044, Japan

**S1. Device fabrication**

The samples used in this work are hexagonal boron nitride (hBN)-encapsulated graphene field effect transistors with an antidot etched inside the channel. The electrical contacts follow a two-terminal geometry with sharp point contacts that protrude toward the circumference of the antidot. The antidot is well isolated from the bulk edges by several microns so that cross-talk from the bulk edges as well as the formation of hot-spots can be neglected. Graphene was encapsulated with hBN using the standard dry-transfer method. A Poly(methyl methacrylate) PMMA mask and E-beam lithography were used to define the anti-dot, which was then etched at 60W, 40/4 sccm $CHF_3/O_2$. The same procedure was used to define the side-contact electrodes and a Ta(5nm)/Nb(40nm) bilayer was then deposited using RF magnetron sputtering at a base pressure of $3 \times 10^{-7}$ Torr. The Ta/Nb bilayer was used for its low contact resistance and the superior step coverage of sputtering[22]. While Nb is a superconductor, in this work we focus on magnetic field above 5T which is significantly above its Hc2 and therefore the effects of superconductivity are of no consequence. Development was done in a low temperature (-15°C) solution of MIBK/IPA(1:3) in order to achieve a sub-30nm resolution. Measurements were carried out in an Oxford Instruments VTI with a 14T superconducting magnet and a He3 refrigerator insert, using the standard lock-in method. A bank of room-temperature pi-filters and low temperature 2-stage RC filters were used to insure the electron temperature closely followed the bath temperature. Excitations used varied from 2.5nA to 10nA at frequencies between 17Hz and 97Hz. Measurements were done at a variety of experimental parameters, including the sweeping rate of gate voltage and magnetic field, in order to confirm the reproducibility of the conductance oscillations.

**S2 Zero Field Device Characteristics**

The mobility for the samples measured can be estimated from the zero field gating curves (Figure S1) using $\mu = \sigma/ne$, where $\sigma$ is the conductivity and $n$ is the carrier density calculated from the geometrical gate capacitance. For this work, the mobility ranged from 28,000 to 50,000

cm$^2$/Vs. The mobility values achieved in these samples are larger than the typical graphene samples fabricated on SiO$_2$ substrate, but can be further improved significantly when compared to the state-of-the-art hBN-encapsulated graphene samples. It was found that the main factors limiting mobility are the remote charge disorder from the SiO$_2$ substrate and contamination on the top hBN surface. Strategies to improve mobility were discussed at length in [1].

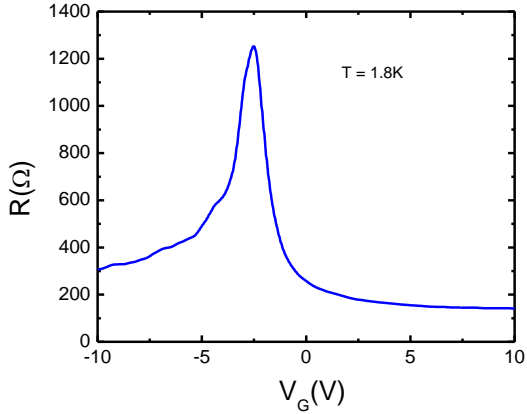

*Figure S1. Typical gating curve for devices measured in this work.*

## S3. Antidot Coupling

Alignments between the antidot and the point contacts were carefully carried out. Due to the resolution and stability limitation of our SEM, there are inevitable variations in the coupling distance/strength between the contacts and the antidot, hence variations in the basic device characteristics. Figure S2 shows the evolution of the gating curve and tunneling amplitude as the distance between the antidot and the leads decreases. The two middle figures are described in detail in the main text, so here we will focus on the two extremes where the contacts are several magnetic lengths away (top) and "cut into" the antidot (bottom). With fully invasive contacts, that is where the leads physically pass through the antidot edge, the resulting QH plateaus have double the usual conductance as would be expected when measuring two, two-terminal devices in parallel. In this case, the phase coherence of electrons in the antidot is destroyed by the metallic leads. It is certain that this sample is monolayer because it is on the same flake as the 200nm device in the main text.

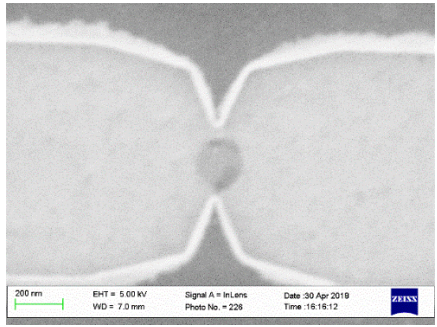 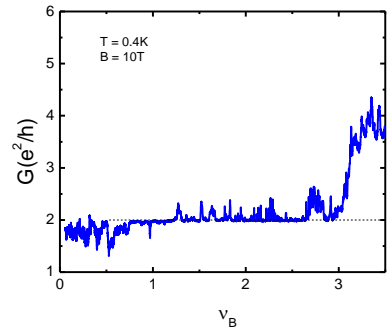

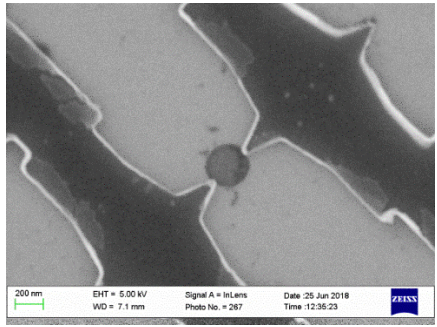 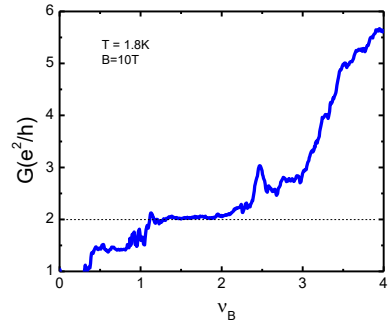

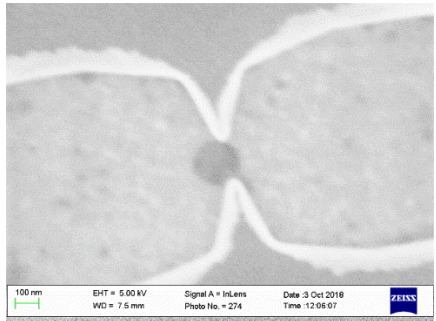 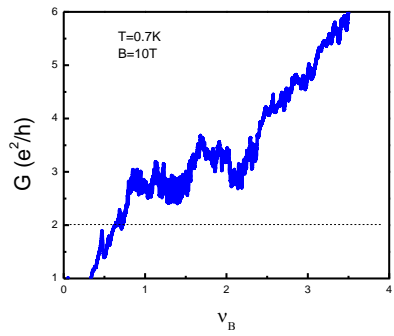

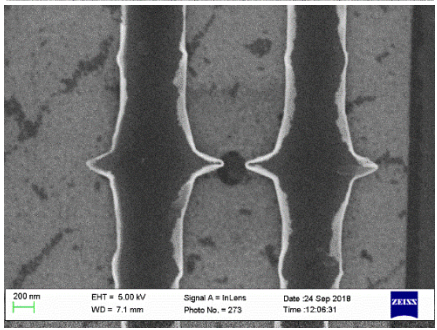 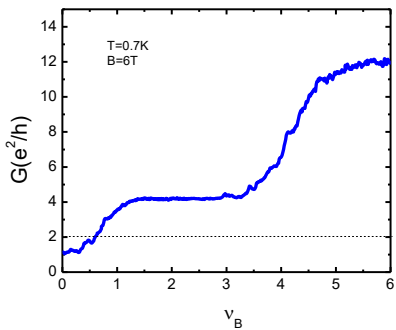

*Figure S2. QH antidots with different coupling strength. 310nm antidot (2$^{nd}$ from the top) and the 200nm antidot (3$^{rd}$ from the top) are discussed in detail in the main text.*

In the opposite limit shown in Figures S2 (top) and S3, where the leads are several magnetic lengths away from the ant-dot edge, there appear randomly spaced oscillations in conductance on a background of the usual conductance value. This is possibly due to the influence of impurities on the tunneling probability.

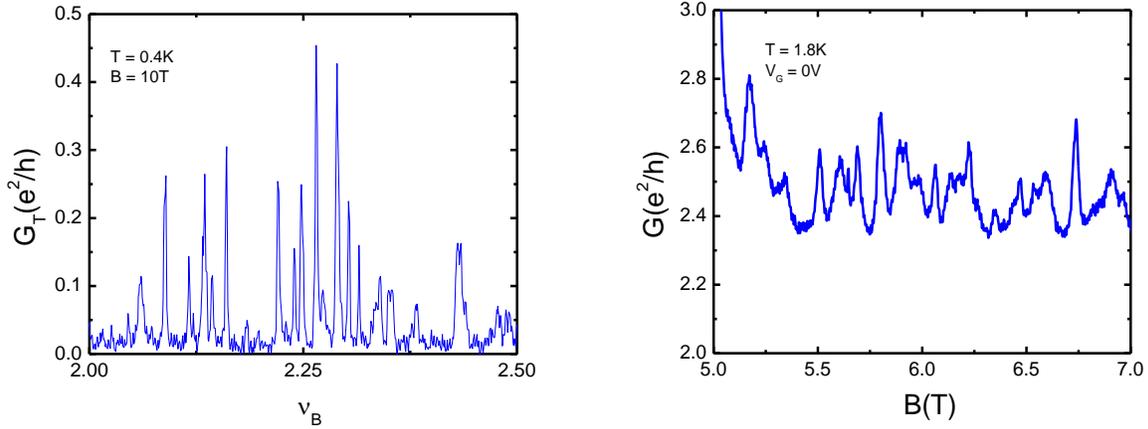

Figure S3. A 200nm antidot sample with contacts around 3 magnetic lengths away from the antidot.

The histogram for gate voltage separation of neighboring conductance peaks is centered on 41 mV, which is the value one would expect given the antidot's 200 nm diameter. The magnetic field separation shows peaks at roughly integer multiples of $\phi_0/2$.

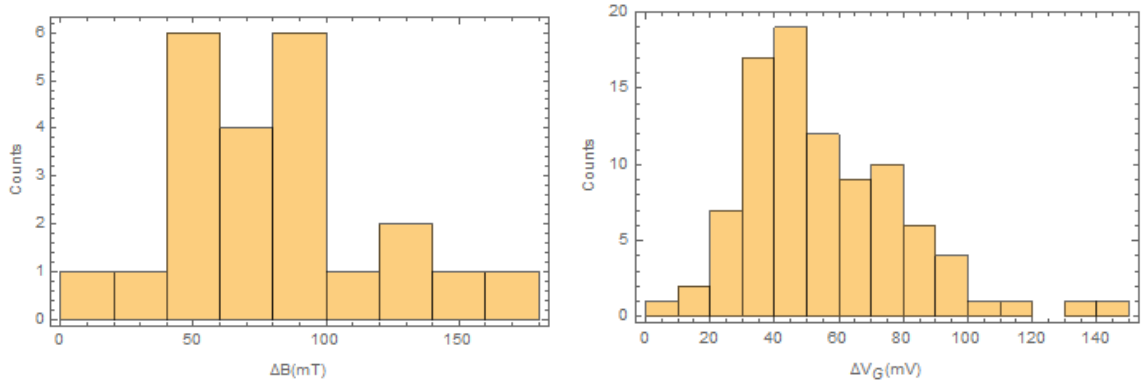

Figure S4. Histograms of flux and gate voltage period for the 200nm sample discussed above. The gating curve distribution on the right is peaked around 41 mV which corresponds to the geometric size of the antidot.

**S4 Edge State Coupling**

In GaAs antidots, the boundary of the antidot was defined either through chemical etching or through depletion using a top gate. The electrons encircling the antidot were then confined to within a depletion length of the physical boundary. Then, again using either a top gate or chemical etching, the edge states of a typical four terminal Hall measurement setup were extended towards the center near the antidot boundary. The distance to the antidot could then be fine-tuned using top gates or front gates. We attempted to mimic this setup by etching a hole in the center of a narrow constriction with a two terminal geometry, in order to demonstrate the feasibility of coupling multiple antidots together by proximity alone (see Figure S5). In graphene, there is no immediately obvious mechanism with which to change the distance between neighboring antidots that is viable at the required length scales (sub 20nm), so the device geometry will be fixed after lithography. With a separation of around 35 nm from the edge states of the constriction to the antidot, oscillations in conductance we observed in magnetic fields between 1.5 and 3 Tesla (see Figure S6). This is consistent with the expectation that the separation should be close to one magnetic length in order to have sufficiently large tunneling probability. This also demonstrates that in high magnetic fields, it will be necessary to have multiple antidots separated by less than 10 nm for there to be adequate coupling.

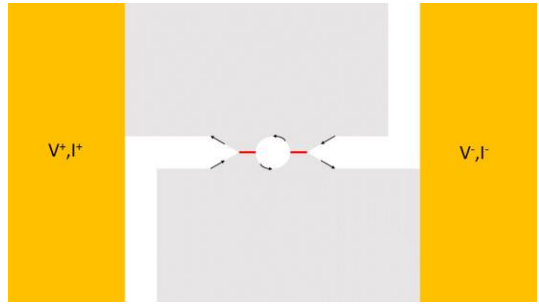

*Figure S5. The sample considered is a two-terminal constriction with a hole etched in the center. The black arrows show the path of the edge states, and the red lines indicate tunneling paths between the edge states and the antidot. The grey area is graphene, the gold area is the gold contacts, and the white area is graphene that has been etched. In zero field, the device shows quantized conductance plateaus at multiples of $8e^2/h$ as one would expect for a constriction carrying two modes.*

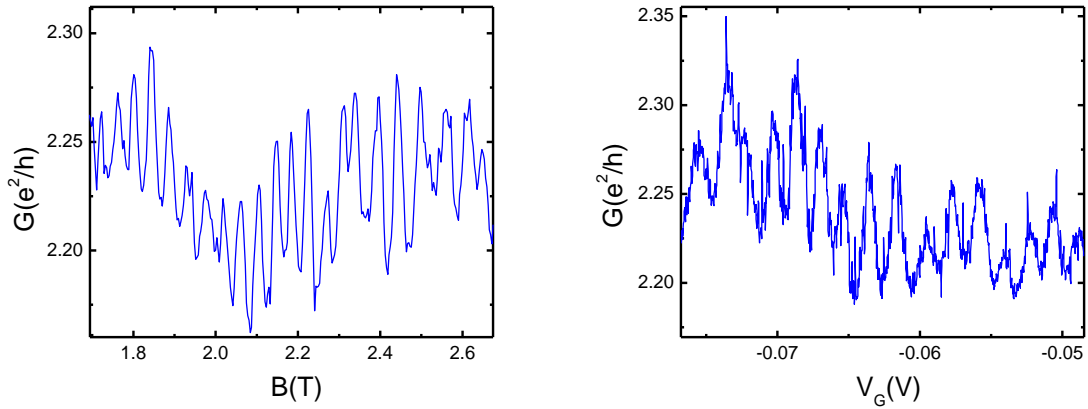

*Figure S6. Edge tunneling into a 250 nm antidot at a bath temperature of around 700mK for the $v = 2$ plateau. In this sample a graphite gate was used and the hBN separating the sample and the back-gate was 15nm.*

## S5 Model Hamiltonian and Aharonov-Bohm Periodicity

The antidot geometry implies that it is natural to write the Schrödinger equation for electrons in graphene in the presence of magnetic field $B$ perpendicular to the graphene plane in the polar coordinates $r, \varphi$ and employing a symmetric gauge for the vector potential $\vec{A} = rB\hat{\varphi}$. The Hamiltonian $H_0$ for electrons in the $K$ valley is (see, e.g. [2])

$$H_0 = v_F \vec{\sigma} \cdot (\vec{p} - e\vec{A}) + U, \quad (1)$$

where $v_F$ is the band-structure velocity near the Dirac point, and $U$ is the potential energy which we can approximate to be constant. For the wavefunction with momentum components close to the Dirac point, the Schrödinger equation in the polar coordinates is

$$\frac{\hbar v_F}{i} \begin{pmatrix} 0 & e^{-i\varphi}(\partial_r - \frac{i}{r}\partial_\varphi + \frac{eBr}{2\hbar}) \\ e^{i\varphi}(\partial_r + \frac{i}{r}\partial_\varphi - \frac{eBr}{2\hbar}) & 0 \end{pmatrix} \begin{pmatrix} u(r,\varphi) \\ v(r,\varphi) \end{pmatrix} = (E-U) \begin{pmatrix} u(r,\varphi) \\ v(r,\varphi) \end{pmatrix}, \quad (2)$$

and $u, v$ are the wavefunction amplitudes on the two graphene sublattices [3], while $E$ is the total electron energy. Assuming rotational symmetry, one can see immediately that the angular part of this equation is satisfied by the solution in the usual angular momentum-conserving form

$$u(r, \varphi) = u_n(r)e^{-i(n+1)\varphi}, \quad v(r, \varphi) = v_n(r)e^{-in\varphi}.$$

The equations for the radial part of the wavefunction then are

$$u'_n(r) + \frac{n+1}{r}u_n(r) - \frac{r}{2l_B^2}u_n(r) = -\frac{E-U}{\hbar v_F}v_n(r) \quad (3)$$

$$v'_n(r) - \frac{n}{r}v_n(r) + \frac{r}{2l_B^2}v_n(r) = \frac{E-U}{\hbar v_F}u_n(r)$$

where we introduced the magnetic length $l_B = \sqrt{\hbar/eB}$.

We are interested in the lowest Landau level (LLL), for which $E - U = 0$, and the only solution of the equation for $u_n$ which satisfies the boundary condition $u_n \to 0$ at infinity is $u_n \equiv 0$. This means that the LLL state in the vicinity of the Dirac point of the $K$ valley are localized on one graphene sublattice. Nonvanishing normalized solutions $v_n$ on this sublattice are:

$$v_n(r) = \frac{1}{(2l_B^2)^{n+1}\sqrt{\pi n!}} r^n e^{-r^2/4l_B^2} \quad (4)$$

These functions are equivalent to the LLL wavefunctions of electrons with nonrelativistic dispersion relation (see, e.g., [4]). For the full graphene plane, the condition that the wavefunction is regular at $r \to 0$ imposes the condition $n \geq 0$ on the index of the wavefunctions. In the presence of an antidot, which makes the point $r = 0$ unaccessible, this restriction does not exist, but the relevant states have $n \gg 1$, and the states with $n < 0$ would be localized very closely to the edge of the antidot and are shifted in energy beyond the relevant energy range (cf. the discussion below).

In the $K'$ valley, the LLL states have the same sublattice structure: they vanish on one sublattice and are given by Eq. (4) on the other, but the sublattices are switched in comparison to the $K$ valley discussed above. This difference makes the response of these states to the boundary conditions at the graphene edge very different.

For the antidots with the diameter $D_{QH}$ on the order of 100 nanometers, e.g., $D_{QH} \simeq 300$ nm, as in the most of the measurements reported in this work, and field $B$ on the order of several tesla, three characteristic distance scales are very different: the size of the antidot, the magnetic length $l_B \sim 10$ nm ($l_B \simeq 11$ nm for $B = 5$ T), and the nearest-neighbor distance $a = 0.14$ nm in the graphene lattice. The continuous description based on Eqs. (2) to (4) is justified by $l_B \gg a$. Also, condition $D_{QH} \gg l_B$ makes it possible to simplify the wavefunctions (4) further. Expanding the logarithm of the right-hand-side of Eq. (4) near the maximum $r_n = \sqrt{2n}\, l_B$, and keeping only the second-order terms, one sees, that in the region where it has significant amplitude, the wavefunctions (4) is Gaussian:

$$\psi_n(y) = \frac{1}{\pi^{1/4} l_B^{1/2}} e^{-y^2/2 l_B^2}, \quad y \equiv r - r_n \quad (5)$$

Equation (5) gives the radial structure of the LLL wavefunctions at distances larger than $l_B$ away from the antidot edge, which on the large scale of the antidot size can be thought of as a circle of radius $D_{QH}/2 \equiv r_e$. Precise structure of the wavefunctions closer to $r_e$ is determined by the boundary conditions at the edge. Circular geometry of the antidot implies that the nature of the boundary condition on the microscopic scale of the graphene lattice changes with the position on the edge. We adopt here the ``zigzag'' boundary conditions which are known to describe most of the generic edges of a graphene lattice [5]. This condition consists in vanishing wavefunction amplitude on one of the graphene sublattices, and is automatically satisfied for LLL states near one of the Dirac points. This means that these states are not affected at all by the edge, and remain at zero energy (see, e.g., [6]) i.e., well below the energy level relevant for electron transport through the antidot. The LLL states at the other Dirac point are non-zero on the sublattice wavefunctions on which should vanish at the edge, and therefore are pushed up in energy by the boundary condition forming one band of the spin-degenerate states that transport the current along the edge and are responsible for the proper Hall conductance $2\, e^2/h$ of the LLL in graphene. In the case of the antidot, this means that we are imposing the boundary condition

$$\psi_n(y = r_e - r_n) = 0$$

on the wavefunctions (5), which become distorted by this condition to various degree depending on how small is the distance to the edge $r_e - r_n$ relative to the magnetic length $l_B$. Under the realistic assumption that the antidot energies are small in comparison to the energy separation between the Landau levels, the LLL states forming the antidot edge states should not be too close to the graphene edge $r_e - r_n \geq l_B$, and the wavefunctions are not modified strongly in comparison to their bulk shape (4). Still, the energy is increased from zero, so that the dispersion relation of the edge state encircling the antidot acquires some finite slope $\hbar v$ that corresponds to the drift velocity of electrons $v \sim v_F$ (see, e.g., [6]) along the antidot edge.

This means that introducing the coordinate $x = r_e \varphi$ along the antidot edge, one can write the coordinate part of the spin-degenerate wavefunctions of the antidot edge states as

$$\psi_j(x,y) = \frac{1}{\sqrt{L}} e^{ik_j x} \psi_j(y), \quad k_j = \frac{2\pi j}{L} = \frac{j}{r_e}, \quad (6)$$

and in the ideal case without disorder, the energies of these states would form an equidistant spectrum

$$\epsilon_j = j\hbar v/r_e + const. \quad (7)$$

For antidots in this work with $r_e \simeq 150$ nm, the energy spacing $\delta\epsilon = \epsilon_j - \epsilon_{j-1}$ in this single-particle spectrum can be estimates as $\delta\epsilon \sim 10$ K taking into account that the velocity $v$ should be somewhat smaller than the band velocity $v_F$. Since the electron $g$-factor in graphene should be close to its free value, the Zeeman splitting of the spin states in the typical magnetic field $B$ of several tesla, $2\mu_B B \sim 5$ K.

So far, the discussion did not include electron-electron interaction. The mean-field effect of the average self-consistent potential energy $U(r)$ created by external electrodes and conducting electrons in graphene itself is simply to change the drift velocity $v$ of the edge states. Indeed, the ``size'' of the states (6) in the radial direction is much smaller than the characteristic scale of the variation of such a potential, which is given by the antidot size, $l_B \ll D_{QH}$. In this case, the only effect of the potential energy on the edge states is to add a constant value of this energy $U(r_j)$ at the location of the state to its energy $\epsilon_j$ (7). Such an addition effectively changes the energy spacing $\delta\epsilon$ and the velocity $v$.

Besides the mean-field effect, electron-electron interaction creates correlations among electrons similar to ``Coulomb-blockade'' correlations in small tunnel structures [7]. Indeed, as follows from the form of edge-state wavefunctions given by Eqs. (5) and (6) the nearest-neighbor states in energy are also the nearest neighbors in the real space, and the real-space distance between them in the radial direction is very small, $\delta r = r_j - r_{j-1} \simeq l_B/\sqrt{2j} \simeq l_B^2/r_e \ll l_B$ in comparison to the magnetic length $l_B$ which determines their size. This means that on the scale of $l_B$ one can neglect $\delta r$, i.e., assume that the few edge states which participate in electron transport through the antidot, are located at essentially the same radius. As a result, the density-density interaction of electrons in these states can be correctly described with one interaction constant $U_{ee}$. This feature makes the antidot transport similar to the Coulomb blockade transport, despite the fact that the origin of this interaction constant in the case of the antidot is very different from both the metallic tunnel junctions and quantum dots. Explicitly, $U_{ee}$ can be evaluated as

$$U_{ee} = \frac{e^2}{4\pi\varepsilon\varepsilon_0} \int d\bar{r} d\bar{r}' |\psi_j(\bar{r})|^2 |\psi_{j'}(\bar{r}')|^2 \frac{1}{|\bar{r} - \bar{r}'|}, \quad (8)$$

where $\bar{r} = \{x,y\}$, and the condition $l_B \ll L$, with $L = \pi D_{QH}$, made it possible to neglect the circular geometry of the antidot states. For the wavefunctions given by Eqs. (5) and (6), this integral can be evaluated with logarithmic accuracy and gives

$$U_{ee} = \frac{e^2}{2\pi\varepsilon\varepsilon_0 L} \ln(L/l_B). \qquad (9)$$

For the typical antidot size we are discussing, $U_{ee} \sim 30$ K. With constant interaction $U_{ee}$, the antidot Hamiltonian has the form

$$H = \frac{U_{ee}}{2} n(n-1) + \sum_{j,\sigma} \epsilon_{j,\sigma} n_{j,\sigma}, \quad n = \sum_{j,\sigma} n_{j,\sigma},$$

where $n_{j,\sigma}$ are the occupation numbers of the edge states with orbital index $j$ and spin $\sigma$, and the single-particle energy $\epsilon_{j,\sigma}$. The total number of electrons $n$ is defined in some strip around the antidot. The precise width of the strip does not affect the results, since varying it only adds a constant to the ``charge'' $n$. As seen below, this change can be absorbed in the definition of zero of the gate voltage, something we cannot keep track of anyway.

The fact that the electron states relevant for tunneling through the antidot essentially overlap in space (as discussed above in the calculation of the constant $U$ of electron-electron interaction for electrons on these states) implies also that the electrostatic coupling constant $\lambda$ of these states to the back-gate voltage $V_g$ is the same for all relevant states. This means that the back-gate voltage shifts uniformly the part of the antidot energy spectrum relevant for transport, and this shift can be separated from the energies $\epsilon_{j,\sigma}$:

$$\sum_{j,\sigma} \epsilon_{j,\sigma} n_{j,\sigma} \to \sum_{j,\sigma} \epsilon_{j,\sigma} n_{j,\sigma} - e\lambda V_g n.$$

Up to an irrelevant overall constant and the shift of $V_g$, the Hamiltonian $H$ above can then be written as

$$H = \frac{U_{ee}}{2}(n - n_g)^2 + \sum_{j,\sigma} \epsilon_{j,\sigma}. \qquad (10)$$

Here $n_g = V_g/\Delta V_g$ is the gate voltage normalized to the period $\Delta V_g = U_{ee}/e\lambda$ with which the backgate voltage modulates the antidot conductance in the ``Coulomb blockade'' regime when the interaction $U_{ee}$ is the dominant energy. Taking into account that the antidot radius $r_e$ is much smaller than the distance $d$ to the back gate, one can model the electrostatics of the antidot as hole of this radius in the conducting plane in the perpendicular electric field $V_g/d$ inside the parallel-plate capacitor formed by graphene and the back gate. Electrostatics of this model can be solved exactly in the oblate spheroidal coordinates and shows that the charge induced around the hole by the gate voltage coincides with the charge induced in the regular parallel plate capacitor (without the hole) in the area $A$ equal to the antidot area, $A = \pi r_e^2$. This gives the estimate of the gate-voltage period $\Delta V_g = ed/A\varepsilon\varepsilon_0$ which is in general agreement with the period observed experimentally.

Interaction of the antidot with another control field, the magnetic flux $\Phi$ through the antidot area is determined by the two effects both of which stem from from the dependence of the LLL states (4) on the magnetic field $B$ through the magnetic length $l_B$. One effect is the increase of the energies (7) of the antidot edge states with increasing magnetic field which can be expressed in

the quasiclassical terms as the interaction between the electron current in the state and the flux $\Phi$:

$$\epsilon_j = j \frac{\hbar v}{r_e} + \frac{ev}{2\pi r_e} \Phi.$$

This interaction gives the proper account of the Aharonov-Bohm effect, which in the case of the antidot manifest itself as the periodicity of the antidot energy spectrum in $\Phi$ with the period equal to the magnetic flux quantum, $\Phi_0$:

$$\epsilon_j(\Phi) = \epsilon_{j-1}(\Phi + \Phi_0). \qquad (11)$$

Another effect that couples the flux $\Phi$ to the antidot dynamics is the charge flow towards the antidot with increasing $\Phi$ through the lower boundary of the strip that defines the charge $n$ on the antidot. For the wavefunctions (4), the shift (11) of the spectrum in the energy space is associated with the similar shift of the wavefunctions in the real space, by one state per flux quantum. To see this explicitly, one can notice that the effective radius $r_j$ of the state $\psi_j(r)$ (5) corresponds to the condition that the orbit area encloses $j$ flux quanta: $B\pi r_j^2 = 2\pi \hbar j/e = j\Phi_0$, and therefore, the states shift by one $\psi_{j-1}(r) \to \psi_j(r)$ upon increase of the magnetic field in such a way that the flux $\Phi$ through these states increases by $\Phi_0$. Since the number $j$ of the flux quanta through the states encircling the antidot is very large, $j \gg 1$, this condition defines the same change of $B$ for all states relevant for transport. The shift of the filled LLL states by one means that the antidot charge $n$, more precisely, the number of electrons $n$ in the strip of fixed width that determines the interaction energy of the antidot, is increased by one per filled LLL when the flux $\Phi$ changes by $\Phi_0$. Therefore, $n$ depends on the flux $\Phi$:

$$n(\Phi) = n + n_\phi, \quad n_\phi = \nu_s \Phi/\Phi_0, \qquad (12)$$

($\nu_s$ being the number of occupied LLs including spin degeneracy which gives the number of the propagating edge modes) and to account for this dependence, the antidot Hamiltonian (10) can be written as follows, up to an irrelevant constant:

$$H = \frac{U_{ee}}{2}\left(n + n_\phi - n_g\right)^2 + \sum_{j,\sigma} \epsilon_{j,\sigma}. \qquad (13)$$

The dependence of the charge $n$ (12) on the flux $\Phi$ through the antidot and the single particle energy spectrum (11) mean that in the relevant regime of the chemical potential fixed by the external contacts to the antidot, the Hamiltonian (13) as a whole is periodic in flux with the period $\Phi_0$ in agreement with the general ``Byers-Yang theorem'' [8].

**S6 IV Curves and Thermal Broadening**

Next, we outline the calculation of the tunnel conductance for tunneling through the antidot between the left and the right metallic electrodes of the structure, governed by the Hamiltonian (13). We do this in the most basic regime, when the spacing $\delta\epsilon$ of the antidot energy spectrum is negligible for some reason, e.g., because of the temperature $T$ that is large on the scale of $\delta\epsilon$. In this case, the charge tunneling is described by the standard ``Coulomb blockade'' rate equation (see, e.g., [7]) for the probabilities $p_n$ of the different charge states of the antidot:

$$\dot{p}_n = J_{n+1} - J_n, \quad J_n = \Gamma_n^- p_n - \Gamma_{n-1}^+ p_{n-1},$$

where the charge tunneling rates $\Gamma$ are the sums over tunneling rates $\Gamma_{L,R}$ in the left and right contacts. The currents in the structure can be expressed in terms of the tunneling rates:

$$I_L = e \sum_n (\Gamma_{L,n-1}^+ p_{n-1} - \Gamma_{L,n}^- p_n),$$

with a similar expression for the current $I_R$ in the right contact. We are interested in the stationary situation, when $\dot{p}_n = 0$, i.e. $J_n = constant \equiv J$. In this situation, the currents in the two contacts should also be equal, $I_L = I_R \equiv I$, the condition that can expressed as $J = 0$. This condition makes it possible to write, explicitly, the stationary solution of the rate equation:

$$p_{-n} = p_0 \prod_{j=0}^{n-1} \Gamma_{-j}^- / \Gamma_{-j-1}^+, \quad p_n = p_0 \prod_{j=0}^{n-1} \Gamma_j^+ / \Gamma_{j+1}^-, \quad n > 0, \qquad (14)$$

and $p_0$ is obtained then from the normalization condition. In equilibrium, when the bias voltage $V$ between the electrodes vanishes, the tunneling rates satisfy the detailed balance condition,

$$\Gamma_n^+ / \Gamma_{n+1}^- = e^{-(U_{n+1} - U_n)/T}$$

and this procedure confirms the equilibrium charge distribution on the antidot:

$$p_n^{(0)} = \frac{1}{Z} e^{-U_n}, \quad U_n = \frac{U_{ee}}{2}(n + n_\phi - n_g)^2.$$

To find the expression for the linear conductance $G$ of the antidot, $I = GV$, one needs to find the linear-in-V correction to the equilibrium solution of the rate equation, which can be done directly in Eq. (14). In this way, we get

$$G = \frac{e^2}{T} \sum_n p_n^{(0)} \frac{\Gamma_{L,n}^+ \Gamma_{R,n}^+}{\Gamma_{L,n}^+ + \Gamma_{R,n}^+}. \qquad (15)$$

In the continuous approximation used above, the tunneling rates depend on the energy change $E$ in the tunneling process as

$$\Gamma_{L,R}(E) = \frac{G_{L,R}}{e^2} \frac{E}{e^{E/T} - 1},$$

and the expression for the linear conductance takes the following form:

$$G = \frac{1}{T} \frac{G_L G_R}{G_L + G_R} \sum_n p_n^{(0)} \frac{U_{n+1} - U_n}{e^{(U_{n+1} - U_n)/T} - 1}. \qquad (16)$$

We use Eq. (16) to calculate numerically the linear tunnel conductance of the antidot in various regimes.

In the opposite regime, when the interaction is effectively negligible, the antidot conductance can be calculated directly from the transmission and reflection probabilities $D_j, R_j = 1 - D_j$ of the two contacts, $j = 1,2$, using the standard double-barrier resonant transmission model:

$$G = \frac{e^2}{4hT} \int d\epsilon \frac{D(\epsilon)}{\cosh^2(\epsilon - \mu/2T)}.$$

Here $D(\epsilon)$ is the net electron transmission probability through the antidot between the two contacts that follows from the double-barrier model:

$$D(\epsilon) = \frac{D_1 D_2}{\left|1 + \sqrt{R_1 R_2} \exp(iL\epsilon/\hbar v)\right|^2}.$$

## S6 Electron-Electron Scattering and Particle Lifetime

One more element of the physics of electron transport through the antidot, relevant at temperatures large on the scale of the single-particle level spacing $\delta\epsilon$, is electron relaxation and dephasing due to electron-electron interaction. Qualitatively, this relaxation smears out the single-particle energy levels and makes electron propagation around the antidot incoherent at large temperatures. The relaxation/dephasing rate can be calculated from the wavefunctions (5) with energy spacing given by (7). Since in this work we use this dephasing only in qualitative arguments, we calculate it below in the most basic approximation, neglecting electron spin. For convenience, we let $L = 2\pi r_e$ be the antidot circumference. The dephasing rate $\gamma$ is calculated with the interaction $V(r)$ as a perturbation using Fermi's Golden Rule:

$$\gamma = \frac{2\pi}{\hbar} \sum_{k_2,k_3,k_4} |\langle k_1, k_2 | V | k_3, k_4 \rangle|^2 f(\epsilon_2)(1 - f(\epsilon_3))(1 - f(\epsilon_4))\delta(\epsilon_1 + \epsilon_2 - \epsilon_3 - \epsilon_4).$$

Here $k_i$ and $\epsilon_i$ are electron momenta and energies before (1,2) and after (3,4) the collision, and $f$ is the Fermi distribution. Momentum conservation implies that the matrix element of the interaction is non-vanishing only if $k_4 = k_1 + k_2 - k_3$, and we can omit the corresponding sum. Letting $q = k_3 - k_1$, the matrix element can be expressed as

$$\langle k_1, k_2 | V | k_3, k_4 \rangle = \frac{1}{L} \int dr V(r) e^{iqr} \equiv V(q).$$

Assuming that electron states are already broadened, e.g., by tunneling or self-consistently by this dephasing itself, we approximate the $\delta$-function over energy by density of states $1/\delta\epsilon$. With these transformations, and taking into account that $\epsilon_3 = \epsilon_1 + \hbar v q$ and $\epsilon_4 = \epsilon_2 - \hbar v q$, the expression for $\gamma$ takes the form:

$$\gamma = \frac{L}{\hbar^2 v} \sum_{k_2, q} |V(q)|^2 f(\epsilon_2)(1 - f(\epsilon_1 + \hbar v q))(1 - f(\epsilon_2 - \hbar v q)) .$$

The matrix element $V(q)$ can be evaluated exactly after neglecting the radial shift of successive wave functions, which is a very good approximation given $r_{n+1} - r_n \ll l_B$ for the range of $n$'s considered:

$$V(q) = \frac{1}{L} \int dr e^{iqr} \int dy_1 dy_2 |\psi_n(y_1)|^2 |\psi_{n'}(y_2)|^2 \frac{e^2}{4\pi\varepsilon\varepsilon_0} \frac{1}{\sqrt{(y_1 - y_2)^2 + r^2}} .$$

Integration of the $y$ coordinates gives the modified interaction

$$V(r) = \frac{e^2}{4\pi\varepsilon\varepsilon_0} \frac{1}{\sqrt{2\pi} l_B} e^{r^2/4l_B^2} K_0(r^2/4l_B^2) ,$$

where $K_0$ is the modified Bessel function. In the limit $r \gg 2l_B$, this expression reduces to the original $1/r$ dependence of the potential. The integral over $r$ can be calculated and gives a similar expression for $V(q)$:

$$V(q) = \frac{e^2}{4\pi\varepsilon\varepsilon_0 L} e^{q^2 l_B^2/4} K_0(q^2 l_B^2/4) .$$

The wavevectors $q$ relevant for the dephasing rate are small, $q \sim 1/L$, and we can use the form of the Bessel function at small arguments, $K_0(z) = -\ln z$, to simplify the above expression to

$$V(q) = \frac{e^2}{2\pi\varepsilon\varepsilon_0 L} \ln(q l_B/2) .$$

Since the interaction matrix elements depend on $q$ weakly, through the logarithm, we can neglect this dependence when evaluating the sums over momenta for the relaxation rate. For the tunneling conductance measurements we are discussing, the relevant relaxation rate $\gamma$ is for electrons at the Fermi energy, i.e. one should take $\epsilon_1 = 0$. In the large-temperature range discussed here, the sums in this equation can be replaced by integrals over energy and give

$$\sum_{k_2, q} f(\epsilon_2)(1 - f(\hbar v q))(1 - f(\epsilon_2 - \hbar v q)) = (\pi k_B T/2\delta\epsilon)^2 .$$

The final expression for $\gamma$ is then

$$\gamma = \frac{L}{\hbar^2 v} \left(\frac{k_B T L}{4\hbar v}\right)^2 \frac{e^4 \ln^2(\pi l_B/L)}{(2\pi\varepsilon\varepsilon_0 L)^2} = \left(\frac{k_B T}{2\hbar}\right)^2 \frac{L}{v} \left(\frac{e^2}{4\pi\varepsilon\varepsilon_0 \hbar v}\right)^2 \ln^2\left(\frac{\pi l_B}{L}\right) .$$